\documentstyle[12pt]{article}
\def\beq{\begin{equation}}
\def\eeq{\end{equation}}
\def\bea{\begin{eqnarray}}
\def\eea{\end{eqnarray}}
\def\nn{\nonumber}
\def\proof{\noindent{\sl Proof.}\ }
\def\psio{{\buildrel \circ \over \psi}}
\def\Psio{{\buildrel \circ \over \Psi}}
\def\tr{{\rm tr}\,}
\def\res{{\rm res}\,}
\def\gr{{\rm grad}_q}
\def\Gr{{\rm Grad}_q}
\def\bone{{\bf 1}}
\def\r#1{{\rm (\ref{#1})}}
\def\c{{\rm can}}
\def\n{\underline{n}}

\def\theor#1#2{\medskip\noindent{\bf Theorem #1.}\  {\it #2}\medskip}
\def\lemm#1#2{\medskip\noindent{\bf Lemma #1.}\  {\it #2}\medskip}
\def\CC{\Bbb{C}}
\def\ZZ{\Bbb{Z}}
\catcode`\@=11
\ifcase\@ptsize
 \font\tenmsy=msbm10
 \font\sevenmsy=msbm7
 \font\fivemsy=msbm5
 \font\teneu=eufm10
 \font\seveneu=eufm7
 \font\fiveeu=eufm5
\or
 \font\tenmsy=msbm10 scaled \magstephalf
 \font\sevenmsy=msbm8
 \font\fivemsy=msbm6
 \font\teneu=eufm10 scaled \magstephalf
 \font\seveneu=eufm8
 \font\fiveeu=eufm6
\or
 \font\tenmsy=msbm10 scaled \magstep1
 \font\sevenmsy=msbm8
 \font\fivemsy=msbm6
\font\teneu=eufm10   scaled \magstep1
\font\seveneu=eufm8
\font\fiveeu=eufm6
\fi
\newfam\msyfam
\textfont\msyfam=\tenmsy  \scriptfont\msyfam=\sevenmsy
  \scriptscriptfont\msyfam=\fivemsy
\def\Bbb{\ifmmode\let\next\Bbb@\else
 \def\next{\errmessage{Use \string\Bbb\space only in math mode}}\fi\next}
\def\Bbb@#1{{\Bbb@@{#1}}}
\def\Bbb@@#1{\fam\msyfam#1}
\newfam\eufam
\textfont\eufam=\teneu  \scriptfont\eufam=\seveneu
  \scriptscriptfont\eufam=\fiveeu
\def\frak{\ifmmode\let\next\frak@\else
 \def\next{\errmessage{Use \string\frak\space only in math mode}}\fi\next}
\def\frak@#1{{\frak@@{#1}}}
\def\frak@@#1{\fam\eufam#1}
\catcode`\@=12
\makeatletter
\newdimen\normalarrayskip              
\newdimen\minarrayskip                 
\normalarrayskip\baselineskip
\minarrayskip\jot
\newif\ifold             \oldtrue            
\def\arraymode{\ifold\relax\else\displaystyle\fi} 
\def\eqnumphantom{\phantom{(\theequation)}}     
\def\@arrayskip{\ifold\baselineskip\z@\lineskip\z@
     \else
     \baselineskip\minarrayskip\lineskip2\minarrayskip\fi}
\def\@arrayclassz{\ifcase \@lastchclass \@acolampacol \or
\@ampacol \or \or \or \@addamp \or
   \@acolampacol \or \@firstampfalse \@acol \fi
\edef\@preamble{\@preamble
  \ifcase \@chnum
     \hfil$\relax\arraymode\@sharp$\hfil
     \or $\relax\arraymode\@sharp$\hfil
     \or \hfil$\relax\arraymode\@sharp$\fi}}
\def\@array[#1]#2{\setbox\@arstrutbox=\hbox{\vrule
     height\arraystretch \ht\strutbox
     depth\arraystretch \dp\strutbox
     width\z@}\@mkpream{#2}\edef\@preamble{\halign \noexpand\@halignto
\bgroup \tabskip\z@ \@arstrut \@preamble \tabskip\z@ \cr}%
\let\@startpbox\@@startpbox \let\@endpbox\@@endpbox
  \if #1t\vtop \else \if#1b\vbox \else \vcenter \fi\fi
  \bgroup \let\par\relax
  \let\@sharp##\let\protect\relax
  \@arrayskip\@preamble}
\def\eqnarray{\stepcounter{equation}%
              \let\@currentlabel=\theequation
              \global\@eqnswtrue
              \global\@eqcnt\z@
              \tabskip\@centering
              \let\\=\@eqncr
              $$%
 \halign to \displaywidth\bgroup
    \eqnumphantom\@eqnsel\hskip\@centering
    $\displaystyle \tabskip\z@ {##}$%
    &\global\@eqcnt\@ne \hskip 2\arraycolsep
         $\displaystyle\arraymode{##}$\hfil
    &\global\@eqcnt\tw@ \hskip 2\arraycolsep
         $\displaystyle\tabskip\z@{##}$\hfil
         \tabskip\@centering
    &{##}\tabskip\z@\cr}
\makeatother
\begin{document}

\hfill DFTUZ/94/23

\hfill {\tt hep-th/9412005}

\bigskip\bigskip
\begin{center}
{\Large  Dressing Technique for Intermediate Hierarchies} \\
\bigskip
\bigskip
{\large P. Holod}\footnote{E-mail: mmtpitp@gluk.apc.org}\\
\bigskip
{\it Institute for Theoretical Physics\\
252143 Kiev, Ukraine}\\
\medskip
{\large and}\\
\medskip
{\large S. Pakuliak}\footnote{E-mail:
pakuliak@cc.unizar.es}
\footnote{On leave of absence from the ITP, Kiev 252143, Ukraine}\\
\bigskip
{\it Departamento de F\'\i sica Te\'orica\\
Facultad de Ciencias\\
Universidad de Zaragoza\\
50009 Zaragoza, Spain}\\
\bigskip
\end{center}
\begin{abstract}
A generalized AKNS systems
introduced and discussed recently in  \cite{dGHM} are considered. It
was shown that  the dressing technique both
in matrix pseudo-differential operators and formal series with respect
to the spectral parameter can be developed for these hierarchies.
\end{abstract}

\setcounter{footnote}{0}

\section{Introduction}

It was shown \cite{dGHM} that besides classical
examples of the hierarchies
of the integrable equations corresponding to the principal and homogeneous
pictures
of the basic representation of the affine Lie algebras
it is possible to consider intermediate cases. These
cases are associated with the different choices of a Heisenberg
subalgebra in a loop algebra $\tilde g$  which in turn correspond
to the different choices of conjugacy class in the Weyl group \cite{KaPe}.
The principal and homogeneous pictures correspond to the
classes defined by Coxeter  and identity elements respectively.

The aim of this paper is to show that for these intermediate
hierarchies the dressing technique in terms of matrix
pseudo-differential operators and formal series with respect to spectral
parameter can be developed  as it was done for
usual generalized KdV$_n$ hierarchies (see \cite{DS}
and reference therein). The starting point of our
consideration will be a Hamiltonian symmetry
reduction and a construction
of the vector field tangent to the reduced manifold and  related to
the integrable flows of the corresponding hierarchy.
Solving the tangency constraints we will find an explicit expression
for the gradient of the gauge-invariant functional  (\ref{e21a})
which ensures the tangency condition.
We will also prove that dressing leads to integrable hierarchies
defined by this gradient.
These results
were utilized previously in the series of papers \cite{DLPPOS}, where
the  nonlocal partners to the integrable hierarchies associated with
the principal realization of $\widetilde{sl}_2$  were considered.

All results in the paper is formulated for the nontwisted loop algebra
$\widetilde{sl}_n$.
The generalization of this approach to the
hierarchies corresponding to $BCD$ series requires the solution
to the problem of classification
the regular elements in the inequivalent Heisenberg
subalgebras of the corresponding loop algebras. This problem was
considered in \cite{DF94}.

The paper is organized  as follows. In the subsection 1.1 we list the
regular elements of the Heisenberg subalgebras of $\widetilde{sl}_n$ which
we will use in our construction. Section 2 is devoted to the Hamiltonian
reduction and solution the tangency constraints.
In the section 3 dressing by matrix pseudo-differential
operators and formal series with respect to
the spectral parameter is discussed and
connection to the integrable hierarchies obtained by the Hamiltonian
reduction is established.

\subsection{Regular elements of the Heisenberg subalgebras
of $\widehat{sl}_n$}

It was shown in \cite{KaPe} that the basic representations of the
affine Lie algebras of the $ADE$  type can be parametrized by a
set of the representatives of the conjugacy classes of the corresponding
Weyl group $W$.  One can associate  with each representative $w\in W$
the Heisenberg subalgebra $s_w$ of the  untwisted loop algebra
$\tilde g$.
The choice  $w=1$ or $w=\hbox{Coxeter element}$ corresponds to the
homogeneous or principal construction of the basic representation.
In  former case the  homogeneous Heisenberg subalgebra
is $\tilde h=\CC[\lambda,\lambda^{-1}]\otimes h$, where $h$ is Cartan
subalgebra of the underlying  simple Lie algebra $g$. In later case
(for $\tilde g=\widetilde{sl}_n$ Kac-Moody algebra) the principal Heisenberg
subalgebra is the linear span of the elements
\beq\label{a1}
a_{k,m}=\lambda^k\left(
\begin{array}{cccc}
&&&\lambda\\ 1&&&\\ &\ddots&&\\ &&1&
\end{array}                        \right)^m,\quad m,k\in\ZZ,\quad
1\leq m\leq n-1.
\eeq

The explicit construction of the all inequivalent graded
Heisenberg
subalgebras  of the algebra $\widetilde{sl}_n$ was proposed in
\cite{tKvL}.
For this algebra the Weyl group is isomorphic to the
symmetric group $S_n$. The conjugacy classes in $S_n$ and hence
the inequivalent Heisenberg subalgebras of $\widetilde{sl}_n$ are
parametrized by partition of $n$;
$\n=n_1+n_2+\cdots+n_p$, $n_1\geq n_2\geq\cdots\geq n_p\geq1$. The
 Heisenberg subalgebra $s_{\n}$ is  spanned by the
elements
\beq\label{e1}
\Lambda=\left(
\begin{array}{cccc}
d_1\Lambda_{n_1}^{\ell_1}&&&\\ &d_2\Lambda_{n_2}^{\ell_2}&&\\ &&\ddots&\\
&&&d_p\Lambda_{n_p}^{\ell_p}
\end{array}\right), \quad \Lambda_{n_i}=\left(
\begin{array}{cccc}
&&&\lambda\\ 1&&&\\ &\ddots&&\\ &&1&
\end{array}                        \right)                ,
\eeq
where $\Lambda_{n_i}$ are $n_i\times n_i$ matrices, $\ell_i$ are arbitrary
integers and $d_i$ are arbitrary {\it distinct} complex numbers,
$i=1,\ldots,p$.

Examining the   eigenvalues of the matrix $\Lambda$
one can prove
\cite{FHM} that the regular graded elements exist only in those
Heisenberg subalgebras of $\widetilde{sl}_n$ which belong to the special
partitions. Namely, the ``equal block partition'' $n=pr$ and ``equal
block plus singlet partition'' $n=pr+1$ ($n_i=r>1$, $\ell_i=\ell$ is
relative prime to $r$,
$i=1,\ldots,p$). In the case $n=pr$ after reordering of the basis the
element $\Lambda$  can be written in the form
\bea\label{e2}
&\Lambda=\lambda^m\Lambda^\ell_r\otimes  D,\quad D={\rm
 diag}(d_1,d_2,\ldots,d_p),\quad \sum_{i=1}^{p}d_i=0,         \nn \\
&1\leq\ell\leq r-1,\quad y_i\neq 0,\quad y_i^r\neq y_j^r,\quad
i\neq j,\quad m\in\ZZ
 .  \eea
The element $\Lambda$ is of grade $(\ell+mr)$ with respect to the
grading operator $r\lambda{d\over d\lambda} + {\rm ad} H$, where the
diagonal matrix $H$ is given by
\beq\label{grade}
H= {\rm diag}[\underbrace{H_r,H_r,\ldots,H_r}_{p\ {\rm times}}],
\quad {\rm where} \quad
H_r={\rm diag}[j,j-1,\ldots,-j],\quad j={r-1\over2}.  \eeq

As we mentioned it was suggested
in \cite{dGHM} to use the graded regular elements of the
inequivalent Heisenberg subalgebras of the loop algebra $\widetilde{sl}_n$
to obtain the generalizations of the Drinfeld-Sokolov hierarchies
\cite{DS}.  In the following sections   we will consider Hamiltonian
reduction and dressing technique based on the grade 1 regular element
$$
\Lambda= \Lambda_{r,p}=\Delta_r\otimes D=I+\lambda e
$$
 of the equal block
partition. It corresponds to the type-I hierarchies
in terms of the papers \cite{dGHM}. For the integrable hierarchies
associated with the partition $n=pr+1$ see \cite{DF94}.

\section{Hamiltonian Reduction}

In the first paragraph of this section we  will sketch  the main steps
of the Hamiltonian reduction for the system associated with the loop
algebra $\widetilde{sl}_n$. The details can be found elsewhere
(see for example \cite{DS}).

Let $M$ be the space of differential operators
${\cal L}=\partial+J$, where $J=q(x)-\lambda e$ is the special
element of the loop algebra          $\widetilde{sl}_n$,
$\partial$ is derivative with respect to the parameter $x\in S^1$
and the function $q(x)$ defines a map $q(x)\colon S^1\to sl_n$.
Following the arguments of the
papers \cite{DS,StS} one can show that the space $M$ is Poisson space
with two natural compatible Poisson brackets.
The action of ad$\ H$ induces the decomposition
of $\widetilde{sl}_n=\widetilde{sl}^+_n\oplus\widetilde{sl}^0_n\oplus
\widetilde{sl}^-_n$, where
$\widetilde{sl}^\pm_n$ and $\widetilde{sl}^0_n$ are block upper,
block lower and block diagonal
matrices. Transformation ${\cal L}\to e^s{\cal L}e^{-s}$,
$s\in \widetilde{sl}^+ _n$
preserve both Poisson brackets as well as monodromy invariants of the
operator ${\cal L}$. First step of the reduction is imposing of
the constraint $q^-=-I$. This special choice of the matrix $q$ corresponds
to the fixing of the image of the momentum map induced by
the Hamiltonian group action $e^s{\cal L}e^{-s}$. Second
step is a factorization of the constrained space $M^{\rm
con}=\left.
M\right|_{q^-=-I}$ by the adjoint action of the group $S=e^s$
\beq\label{e7}
{\cal L}^{\rm con}\to   S{\cal L}^{\rm con}S^{-1},\quad q-\Lambda\to
S(q-\Lambda)S^{-1}-\partial S\cdot S^{-1}, \eeq
where now   ${\cal L}^{\rm con}=\partial+q-\Lambda_{r,p}$
 and $q\in \widetilde{sl}^+_n\oplus\widetilde{sl}^0_n$.
The action \r{e7} means that the space $M^{\rm con}$ is an orbit of the adjoint
action of the group $S$. It is well known \cite{DS} that
it is possible to separate different
orbits by introducing the $r$-order (matrix in our case)
differential operator $L$
\beq\label{e8}
L=D^{-r}\partial^r+\sum_{i=0}^{r-1}u_i\partial^i,
\eeq
where $u_i$ are $p\times p$ matrices with entries which are differential
polynomials of the elements of the matrix $q$.
To obtain exact expression for the operator $L$ we have to consider
linear problem ${\cal L}\Psi=0$ for the column $\Psi$ and exclude first $n-p$
components of $\Psi$. The remaining $p$ components  obey an
equation $\sum_{j=n-p+1}^{n}L_{ij}\Psi_j=
\lambda\Psi_i$,  $i=n-p+1,\ldots,n$. The group element $S$
 acts on the column $\Psi$ by left multiplication $\Psi\to S\Psi$, so
 the
 last $p$ components of the column $\Psi$ do not change under
this transformation.
 It means that the  operator $L$ is indeed invariant with respect to
 transformation \r{e7} and hence can be chosen
to parametrize the reduced factor space
 $\overline M=M^{\rm con}/S$.

There exists more formal way to obtain the operator $L$. Following \cite{DS}
let us write  operator ${\cal L}$ in the form
\beq\label{e9}
{\cal L}=\left(
\begin{array}{cc}
\alpha&\beta\\ A&\gamma
\end{array}   \right),
\eeq
where $\alpha$, $\beta$, $\gamma$ and $A$ are the $p\times(n-p)$, $p\times p$,
$(n-p)\times p$ and $(n-p)\times(n-p)$ matrices respectively. Then
$p\times p$ matrix differential operator $L$ can be obtained by the formula
\beq\label{e10}
L=D^{-1}(\beta-\alpha A^{-1}\gamma).
\eeq

Let us fix the gauge for the operator ${\cal L}$. We
choose the following canonical gauge
\beq\label{e11}
{\cal L}^\c=\partial+q^\c-\Lambda_{r,p},\quad{\rm where}\quad
q^\c=\left(\begin{array}{ccccc}
v_{r-1}&v_{r-2}&\cdots&v_1&v_0\\
\\ &&0&&\\ \\
\end{array}\right),
\eeq
and $v_i$ are $p\times p$ matrices. By this choice the relations
between matrices $u_i$ and $v_i$, $i=0,\ldots,r-1$ have
the simplest form
\beq\label{e12}
u_i=D^{-1}v_iD^{-i}.
\eeq

In the sequel we will need only  a second Poisson bracket which in
the space $M^{\rm con}$ can be defined as follows.
For two  functionals $f(q)$ and $g(q)$ associated with the operator
${\cal L}$ define the Poisson bracket
\beq\label{e4}
\{f,g\}(q)=\int_{S^1}\tr(\partial+q-I)[\gr f,\gr g]dx  \ ,
\eeq
where $\gr g$ is given by
\beq\label{e5}
\left.{dg(q+\varepsilon h)\over d\varepsilon}\right|_{\varepsilon=0}=
\tr (\gr g\cdot h).
\eeq

Define the gauge invariant functional on the space $M^{\rm con}$ by
the formula
\beq\label{e13}
\ell_X=\int_{S^1}\tr\res(LX)dx,
\eeq
where $X$ is $p\times p$ matrix  pseudo-differential operator of the form
\beq\label{e14}
X=\sum_{j=1}^{\infty}\partial^{-j}\circ X_j
\eeq
and \res\ means  the coefficient at  $\partial^{-1}$.
Using the formulas \r{e8},\r{e10},\r{e5} and
arguments of the paper \cite{DS} the matrix elements of the
 gradient
of the functional $\ell_X$ read  as
\beq\label{e15}
(\gr\ell_X)_{ij}=\res(D^{i-r}\partial^{r-i}X (L
\partial^{j-r-1})_+D^{r-j}),\quad i,j=1,\ldots, r,
\eeq
where indices $i,j$ numerate the $p\times p$ blocks of the gradient.
Upper triangular blocks
of the $\gr\ell_X$ are not determined by \r{e5} and  Poisson bracket
\r{e4} on
the gauge invariant functionals does not depend on the choice of the
upper triangular part of the $\gr \ell_X$ and are correctly defined.

Using \r{e15} one can easily calculate the vector field
$[\partial+q^\c-I,\gr\ell_X]$. The nonzero entries of this field are
 \bea\label{e16}
&\left([\partial+q^\c-I,\gr\ell_X]\right)_{1j}=
D(LX)_-L\partial^{j-r-1}D^{j-r},\nn\\
&\left([\partial+q^\c-I,\gr\ell_X]\right)_{1r}=
D[(LX)_-L\partial^{-1}-
(\partial D^{-1})^{r-1}(XL)],\nn\\
&\left([\partial+q^\c-I,\gr\ell_X]\right)_{ir}=
-(\partial D^{-1})^{r-i}XL,
\eea
where $j=1,\ldots,r-1$ and $i=2,\ldots,r$.
It is seen from \r{e16} that this vector field is tangent to $M^{\rm con}$
but
is not tangent to reduced manifold $\overline{M}$. The condition of
tangency yields the $r-1$ constraints for the operator $X$. Indeed, the
conditions $\res XL=\res \partial  D^{-1}  XL
= \res (\partial D^{-1})^{r-2}XL=0$, which are equivalent to
\beq\label{e17}
\res XL=\res \partial XL = \res \partial ^{r-2}XL=0,
\eeq
allow one to express matrix coefficients $X_{r+1},\ldots,
X_{2r-1}$ of the operator $X$ via the coefficients
$X_1,\ldots,X_r$ and matrix coefficients of the
operator $L$.

Using the identity \cite{DS,LP}
 \beq\label{e18}
 [L(XL)_-]_+=\sum_{a=1}^{r}(L\partial^{-a})_+\circ \res(\partial^{a-1}XL)
 \eeq
 which is valid for any differential and pseudo-differential operators
 $L$ and $X$  one can
 show that
 if operator $X$ satisfy \r{e17}, than only zero coefficient of the
 differential operator $[L(XL)_-]_+$ is nonzero
 $$
 [L(XL)_-]_0=\res (D^{-r}\partial^{r-1}XL).
 $$
$[B]_i$ means the coefficient at $\partial^i$ of the operator $B$.
Thus the nonzero entries of the
vector field \r{e16} after solution of the constraints
 \r{e17} are
 \beq\label{e19}
\left(\left.[\partial+q^\c-I,\gr\ell_X]\right|_{\rm const=0}\right)_{1j}
=D[(LX)_-L-L(XL)_-]\partial^{j-r-1}D^{j-r},
\eeq
where  $j=1,\ldots,r$.
It is possible to solve the constraints \r{e17} in the
expression for the gradient \r{e15}. Indeed, for $j>i$ we have
\bea
\res(D^{i-r}\partial^{r-i}X (L \partial^{j-r-1})_+D^{r-j})&=&
\res(D^{i-r}\partial^{r-i}X L \partial^{j-r-1}D^{r-j})-\nn\\
&-&\res(D^{i-r}\partial^{r-i}X (L \partial^{j-r-1})_-D^{r-j}), \nn
\eea
and second summand in this formula does not depended on the matrices
$X_{r+1},\ldots,$ $X_{2r-1}$. Using the obvious identity
\bea
\res \partial^{r-i}XL\partial^{j-r-1}&=&
\sum_{a=0}^{r-1} C^{a+j-i}_{a+r-i}
\partial^{a+j-i}\res(\partial^{-a-1}XL)+\nn\\
&+&\sum_{b=1}^{j-i}
C^{j-i-b}_{r-i-b}\partial^{j-i-b}\res(\partial^{b-1}XL)\nn
\eea
the expression for the $\left(\left.\gr\ell_X\right|_{{\rm constr}=0}\right)
_{ij}=\left(\Gr\ell_X\right)_{ij}$ is
\beq\label{e21a}
\left(\Gr\ell_X\right)_{ij}=\left\{
\begin{array}{ll}
\res(D^{i-r}\partial^{r-i}X (L \partial^{j-r-1})_+D^{r-j}), &j\leq i\\
-\res(D^{i-r}\partial^{r-i}X (L \partial^{j-r-1})_-D^{r-j}) +& \\
+D^{i-r}\sum_{a=0}^{r-1} C^{a+j-i}_{a+r-i}
\partial^{a+j-i}\res(\partial^{-a-1}XL) D^{r-j}, &j>i,
\end{array}
\right.
\eeq
where $C^b_a$ are binomial coefficients and $1\leq i,j\leq p$ numerate
the blocks of $rp\times rp$ matrix $\Gr\ell_X$.

{}From \r{e19} and \r{e15} follow that Poisson bracket \r{e4} for
the functionals $\ell_X$ and $\ell_Y$ coincides with the second
Gelfand-Dickey bracket \cite{GDprep}
\beq\label{e20}
\{\ell_X,\ell_Y\}=\int_{S^1}\tr\res[(LX)_-LY-YL(XL)_-].  \eeq

The auxiliary linear problem for the operator $L$  is the
linear problem for the $p$-component KdV$_r$ hierarchy studied previously
in \cite{GDFAP,Man,Wil}.\footnote{The requirement ${\rm diag}\,u_{r-1}=0$
has to be imposed in this case on the operator $L$.} In
\cite{FHM} the monodromy invariants for the operator     $L$
were calculated. Let us briefly formulate this result.

Since all entries of the diagonal matrix $D^{-r}$ are  different,
it is possible                     to find the operator of the form
$g=\bone_p+\sum_{i=1}^{\infty}g_i\partial^{-i}$ using the
recursion procedure such that
\beq\label{e21}
\hat L=g^{-1}Lg=D^{-r}\partial^r+\sum_{i=1}^{\infty}a_i\partial^{r-i},
\eeq
where $a_i$ are the diagonal matrices. The matrices
$g_i$ can be uniquely defined
from \r{e21} by requirement that all ${\rm diag}\ g_i=0$
  \cite{DS,FHM}.  The involutive set of the monodromy invariants
for the operator $L$ are
\beq\label{e22}
{\cal H}_{0,i}=\int_{S^1}(D^ru_{r-1})_{ii},\quad {\cal
H}_{k,i}=\int_{S^1}\res\left(\hat L^{k/r}\right)_{ii},
i=1,\ldots,p;\quad k=1,2,\ldots\ ,
\eeq
and fractional power of the operator $\hat L$ is  pseudo-differential
operator with the property $\left(\hat L^{1/r}\right)^r=\hat L$.
In case of $p$-component KdV$_r$ the Hamiltonians ${\cal H}_{0,i}$ are
absent.

\section{Dressing Technique for the $p$-Component KdV$_r$\newline
 Hierarchies}

\subsection{Dressing by matrix pseudo-differential operators}

The aim of this subsection is the formulation of  the Zakharov-Shabat
dressing technique applied
to the $p\times p$ matrix differential operator $L$
of the $r$th order.
We will proceed in a way similar to the standard scalar dressing
procedure for the usual   KdV$_n$ hierarchies.

Let us  consider the operators $L=(\Delta\partial)^r+\sum_{i=0}^{r-1}
u_i\partial^i$, $L_0=(\Delta\partial)^r$ and
columns  $\psi$, $\psio$ which are the solutions to the
corresponding linear problems
\beq\label{e23}
L\psi=\lambda\psi,\quad L_0\psio=\lambda\psio,\quad\Delta=D^{-1}={\rm diag}\,
(\Delta_1,\ldots,\Delta_p).
\eeq
Suppose that solutions $\psi$ and $\psio$  are connected by
invertible matrix
valued operator Volterra $K(x,\partial)$
\beq\label{e24}
\psi=K(x,\partial)
\psio=\left(\bone_p+\sum_{j=1}^{\infty}K_j(x)\partial^{-j}\right) \psio.
\eeq
Equation \r{e23} yields  the recurrence relation for the matrix
coefficients of the operator $K$
\bea        \label{e27}
[K_{i+r},\Delta^r]&=&\sum_{k=0}^{r-1}{1\over
k!}\left({\partial^kL\over\partial\ \partial^k}\right)\cdot
K_{i+k},\nn\\
i&=&1-r,2-r,\ldots,\quad K_0=1,\quad K_j=0,\quad j<0,
 \eea
 where $\left({\partial^kL/\partial\ \partial^k}\right)$ means the
 $k$-fold formal derivative of the differential operator $L$ with
 respect to the symbol $\partial$ and $(B)\cdot K_i(x)$ means the action
 of the differential operator $B$ applied
to the function $K_i(x)$. For example,
 first two equations from the set \r{e27} are
\beq
[K_1,\Delta^r]=u_{r-1},\label{e25}\eeq
\beq
[K_{2},\Delta^r]=r\Delta^r
K'_{1}+u_{r-1}K_{1}+u_{r-2},\label{e26}\eeq
$K'_i$ means the
derivative with respect to $x$.

\lemm{1}{If ${\rm diag}\,u_{r-1}=0$ the recurrence relations \r{e27}
for the matrix functions $K_i$ can be solved.}

\proof\  Since the entries $\Delta_i^r$ of the diagonal matrix $\Delta^r$
are all different it is easy to see that diagonal part of the equation
\r{e27} determines the diagonal elements of the matrix
$(K_{i+r-1})_{jj}$, while the off-diagonal part of the same equation
yields the off-diagonal elements of the matrix $(K_{i+r})_{jk}$, $j\neq
k$. For example, the off-diagonal elements of the matrix $K_1$ defined
by \r{e25} and the diagonal elements of the same matrix can be
obtained from \r{e26}
\bea\label{e28}
(K_1)_{kj}&=&{(u_{r-1})_{kj}\over\Delta^r_j-\Delta^r_k},\quad k\neq
j,\nn\\
r\Delta^r_i(K_1)'_{jj}&=&\sum_{k=1}^{p}{(u_{r-1})_{jk}(u_{r-1})_{kj}\over
\Delta^r_k-\Delta^r_j}-(u_{r-2})_{jj},\quad j=1,\ldots,p.
\eea

To define the time evolution of the operator $L$ we introduce
$p\times p$ matrices
$E_a={\rm diag}\,(0,\ldots,0,1,0,\ldots,0)$ with 1 on the $a$th place
and infinite set of the time variables $t_s^{(a)}$, $a=1,\ldots,p$,
$s=1,2,\ldots\ $ .  Let us define the time evolution of the
operator $K$ and the solution $\psio$ of the initial linear problem
by
\beq\label{e29}
{\partial K\over\partial
t_s^{(a)}}K^{-1}+(K(\Delta\partial)^sE_aK^{-1})_-=0,\quad
 {\partial\psio\over\partial t_s^{(a)}}=(\Delta\partial)^sE_a\psio.
\eeq
The following theorem is valid.

\theor{2}{The compatibility condition for the linear problems $L\psi=\lambda
\psi$ and $\partial_{t_s^{(a)}}\psi=(K(\Delta\partial)^sE_aK^{-1})_+\psi$
\beq\label{e29a}
{\partial L\over\partial t_s^{(a)}}=(LX)_+L-L(XL)_+,\quad {\rm where}\quad
X=(K(\Delta\partial)^{s-r}E_a K^{-1})_-
\eeq
is
equivalent to the evolution equation for the matrix $q^\c$ defined by
the bracket \r{e4} with Hamiltonian ${\cal H}_{s,a}$ from the set
\r{e22}.}

\noindent
The proving of this theorem is based on the  following Lemma

\lemm{3}{The  operator $A=g^{-1}K$, where operators $K$ and $g$ defined by
\r{e27} and \r{e21} respectively, is diagonal pseudo differential operator.}

\proof\ Suppose that operator $A$ has nonzero off-diagonal part and
 define the column vector $\hat\psi=g^{-1}\psi=A\psio$.
It satisfies the linear problem $\hat
L\hat\psi=\lambda\hat\psi$, where $\hat L$ is diagonal operator
defined by \r{e21}. Now  the relations $\hat LA=AL_0$ and $\Delta_i^r
\not=\Delta_j^r$, $i\not= j$ yield that all off-diagonal
entries of the operator $A$ are identically zero.

The statement of the theorem 2 follows now from the equation
\r{e19} and  the    relation
\bea\label{e29b}
\ell_X&=&\int_{S^1}\tr\res(LX)dx\nn\\
&=&\int_{S^1}\tr\res(K(\Delta\partial)^rK^{-1})
(K(\Delta\partial)^{s-r}E_aK^{-1})_- dx\nn\\
&=&\int_{S^1}\tr\res K(\Delta\partial)^{s}K K^{-1}E_aK^{-1} dx\nn\\
&=&\int_{S^1}\tr\res (g^{-1}K(\Delta\partial)^{s}K^{-1}g)(
g^{-1}K  E_a(g^{-1}K)^{-1})
 dx\nn\\
 &=&\int_{S^1}\tr\res (\hat L^{s/r} E_a)
dx= \int_{S^1}\res \left(\hat L^{s/r} \right)_{aa} dx =
{\cal H}_{s,a}.
\eea
In the second line of \r{e29b} we used the fact that $L$ is
differential operator and the operator $\hat
L^{1/r}=g^{-1}K(\Delta\partial) (g^{-1}K)^{-1}$ obviously satisfies
the relation
$\left(\hat L^{1/r}\right)^r=\hat L$.

It follows from the definition of the time flows that
$\partial/\partial x=\sum_{a=1}^{p}d_a\partial/\partial t_1^{(a)}$,
$d_a=\Delta_a^{-1}$.  Indeed,
\beq\label{e31} \sum_{a=1}^{p} y_a
{\partial\psi\over\partial t_1^{(a)}}=(K\partial K^{-1})_+\psi=
{\partial\psi\over\partial x}.
\eeq
Moreover it can be shown that operator $L$ and hence ${\cal L}^\c$ does
not depend on the certain combination of the times $t_s^{(a)}$.
{}From \r{e22} and \r{e23} follows that
\beq\label{e31a}
\sum_{a=1}^{p}{\cal H}_{mr,a}=0,\quad m=1,2,\ldots\
\eeq
and hence
\beq\label{e31c}
\sum_{a=1}^{p}
{\partial L\over\partial t_{mr}^{(a)}}=0.
\eeq

Except continuum times $t_s^{(a)}$, $s\geq1$ we can introduce $p$
discrete times $t_0^{(a)}$, $a=1,\ldots,p$.
The evolutions of column $\psio$ and operator
$K$ with respect to these times are given
by\footnote{In what follows we will write the dependence on the discrete times
explicitly only if they are changed and expression
$K(t_0^{(a)}+1)$ will mean the shift of discrete time $t_0^{(a)}$
in the coefficients of the operator $K$ by 1
 for fixed $a$.}
\bea\label{e*}
&\psio(t_0^{(a)}+\delta_{ab})=(\Delta\partial)E_b\psio  ,\nn\\
&\left[K(t_0^{(a)}+\delta_{ab}) (\Delta\partial)E_b K^{-1}\right]_-=0,\quad
a,b=1,\ldots,p.
\eea
The compatibility  conditions
of the linear problems
\bea\label{linear}
\psi(t_0^{(a)}+1)&=&\left[K(t_0^{(a)}+1) (\Delta\partial)E_a
K^{-1}\right]_+ \psi=\left[(\Delta\partial)E_a+m_a\right]\psi,\nn\\
{\partial\psi\over\partial t_0^{(b)}} &=&
\left[K(\Delta\partial)^sE_b
K^{-1}\right]_+ \psi ,
\eea
yield the $p$-component analog of the modified KdV$_r$ hierarchy. Indeed, in
the simplest case $p=1$, $r=2$ and $s=3$ this compatibility condition reads
\bea\label{mKdV}
{\partial m\over \partial t_3} &=&
\left[K(t_0+1)\partial^3K(t_0+1)^{-1}\right]_1 m'-
\left[K(t_0)\partial^3K(t_0)^{-1}\right]'_0 \nn\\
&&-m
\left[K(t_0+1)\partial^3K(t_0+1)^{-1}-K(t_0)\partial^3K(t_0)^{-1}\right]_0\nn\\
&=&{1\over4}m'''-{3\over2}m^2m',\nn
\eea
where
$K(t_0+1)$,   $K(t_0)$ are defined by
\bea\label{red}
K(t_0+1)\partial^2K(t_0+1)^{-1}&=&\partial^2+u(t_0+1)=(\partial+m)(\partial-m)
\nn\\
K(t_0)\partial^2K(t_0)^{-1}&=&\partial^2+u(t_0)=(\partial-m)(\partial+m).
\nn
\eea

{}From the discrete evolutions \r{e*} follows that the operator
${\cal K}=\sum_{a=1}^{p}K(t_0^{(a)}+r)E_a$
will satisfies the relation
\beq\label{e**}
\left({\cal K}(\Delta\partial)^rK^{-1}\right)_-=0.
\eeq
This equation can be solved if we impose the restriction
\beq\label{e***}
{\cal K}=\sum_{a=1}^{p}K(t_0^{(a)}+r)E_a=K
\eeq
which is nothing but reduction from $p$-component KP hierarchy to the
$p$-component KdV$_r$ hierarchy expressed in terms of the dressing
operators and this reduction corresponds to the equal block gradation
of the loop algebra  $\widetilde{sl}_n$
described above. In the case of one-component theory the relation
\r{e***} shows that corresponding $\tau$-function is periodic
function of discrete time with period equal to the order of the
differential operator $L$.
The relation \r{e***} yields the generalized Miura
transformation
\bea
&\sum_{a=1}^{p}
\left((\Delta\partial)E_a\!+\!m_a(t_0^{(a)}\!+\!r\!-\!1)\right)\ldots
\left((\Delta\partial)E_a\!+\!m_a(t_0^{(a)}\!+\!1)\right)
\left((\Delta\partial)E_a\!+\!m_a(t_0^{(a)})\right)\nn\\
&\qquad\qquad\qquad\qquad\qquad\qquad=
(\Delta\partial)^r+\sum_{i=0}^{r-1}u_i\partial^i, \label{Miura}
\eea
where
\beq\label{miu}
m_a=K_1(t_0^{(a)}+1)\Delta E_a -E_a\Delta K_1(t_0^{(a)})
\eeq
and $K_1$ is the coefficient at the $\partial^{-1}$ in the operator
$K$.
Generalized Miura transformation \r{Miura} respect the condition
diag$\,(u_{r-1})=0$. Indeed,
\bea\label{e****}
{\rm diag}(u_{r-1})&=&{\rm diag}\sum_{a=1}^{p}
\left[K_1(t_0^{(a)}+r)\Delta^rE_a-E_a\Delta^rK_1(t_0^{(a)})\right.\nn\\
&&+\ldots+\left.K_1(t_0^{(a)}+1)\Delta^rE_a-E_a\Delta^rK_1(t_0^{(a)}+1)\right]
=0\nn
\eea
due to \r{e***} and \r{e25}. The relation \r{e***} means that shifts on $r$
with respect to each discrete times are not independent and this relation
is the discrete counterpart of the relation \r{e31c}.

We will show in the next subsection that the Hamiltonians of the $p$-component
modified KdV$_r$ hierarchy are defined by \r{e22} up to some
gauge transformation with fields $u$ replaced by fields $m$ using \r{Miura}.

\subsection{Dressing by formal series with respect to the spectral parameter}

The goal of
this subsection is to show that $p\times p$ matrix  representation
\r{e29a} in terms of the matrix differential operators of the $r$th
order can be lifted to the $pr\times pr$ matrix representation of the
differential operator of the first order depending on the spectral
parameter $\lambda$.  Our arguments here will be similar to the ones
presented in \cite{DLPPOS}.

We start from the initial linear problem for the $n$-size column $\Psio$
\beq\label{e32}
(\partial-\Lambda)\Psio=0,\quad \Lambda=\Lambda_{p,r}
\eeq
and define the formal $n\times n$ matrix  series in $\lambda$ by the equation
\beq\label{e33}
G(\Lambda)=\sum_{i=0}^{\infty}G_i\Lambda^{-i},\quad
\left(G_i\right)_{ab}=\left\{\begin{array}{ll}
\Delta^{r-a}C^{b-a}_{r-a}\partial_x^{b-a}K_i\Delta^{b-r} &b\geq a,\\
0&b<a,\end{array}\right.
\eeq
where $a,b=1,\ldots,r$ and
$K_i$ are some $p\times p$ matrix function depending on all times
$t^{(a)}_i$.   Let $\Psi=G(\Lambda)\Psio$ and suppose that formal series
$G(\Lambda)$ is subjected to the following constraints
\beq\label{e34}
G(\Lambda)(\partial_x-\Lambda)G^{-1}(\Lambda)=\partial_x-\Lambda+q^\c,
\eeq
\beq\label{e35}
G(\Lambda)(\partial_{t_s^{(a)}}-\Lambda_r^s\otimes E_a)G^{-1}(\Lambda)=
\partial_{t_s^{(a)}}-V(\lambda)=\partial_{t_s^{(a)}}-\sum_{i=0}^{s}
V_{i}\lambda^i.
\eeq
These  constraints show that $\Psi$ is the simultaneous solution to the
following linear problems
\beq\label{lin1}
(\partial_x-\Lambda+q^\c)\Psi=0,
\eeq
\beq\label{lin2}
(\partial_{t_s^{(a)}}-V(\lambda))\Psi=0.
\eeq
The following statement is valid.

\theor{4}{The constraints \r{e34} and \r{e35} for the unknown functions
$K_i$ are equivalent to the
equations \r{e27} and \r{e29} for the matrix Volterra operator
$K=\bone_p+\sum_{i=1}^\infty K_i\partial^i$.
The zero coefficient with
respect to $\lambda$ of the
matrix polynomial        $V(\lambda)$ coincides with the expression \r{e21a}
for the $\Gr {\cal H}_{s,a}$}.

\proof\  The equivalence of \r{e34} and \r{e27} follows from the
straightforward calculations using \r{e32}  and the fact that
$\Delta^r_i\not=\Delta^r_j$, $i\not= j$. To prove the equivalence
of \r{e35} and \r{e29} we need the following technical lemma.

\lemm{5}{If $\Psi=(\psi_1,\ldots,\psi_r)^t$ is the solution to the
equation \r{lin1}, where $\psi_i$ are $p$-component vectors, then
\beq\label{negative}
\partial_x^{-(i+1)}\psi_r=O(\lambda^{-1})\psi_r+
O(\lambda^{-1})\psi_{r-1}+\ldots+O(\lambda^{-1})\psi_1, \quad i\geq0.
\eeq                                                                 }

\proof\  The straightforward calculations similar to the ones presented in
\cite{DLPPOS}.

Now the statement about equivalence of \r{e35} and \r{e29} follows from
the relation
\bea\label{equiv}
&&V(\lambda)\Psi=(G(\Lambda)_{t_s^{(a)}}+G(\Lambda)\Lambda_r^s\otimes E_a)
\Psio=\nn\\
&&\quad=\left(\begin{array}{c}
\Delta^{r-1}\left[(K(\partial)_{t_s^{(a)}}K(\partial)^{-1}+
K(\partial)(\Delta\partial)^sE_aK(\partial)^{-1})\psi_r\right]^{(r-1)}\\
\Delta^{r-2}\left[(K(\partial)_{t_s^{(a)}}K(\partial)^{-1}+
K(\partial)(\Delta\partial)^sE_aK(\partial)^{-1})\psi_r\right]^{(r-2)}\\
\vdots\\
(K(\partial)_{t_s^{(a)}}K(\partial)^{-1}+
K(\partial)(\Delta\partial)^sE_aK(\partial)^{-1})\psi_r
\end{array}\right)
\eea
and Lemma 5. Expression $[f]^{(i)}$ means here the $i$-fold derivative
of the function $f$ with respect to $x$.
The absence of negative powers of $\partial$ in the
operator $(K_{t_s^{(a)}}K^{-1}+
K(\Delta\partial)^sE_aK^{-1})$
means the absence of negative powers of $\lambda$ in the matrix $V(\lambda)$.
Now the zero coefficient of the matrix $V(\lambda)$ with respect to $\lambda$
can be obtained from \r{equiv} using $L\psi_r=K(\Delta\partial)^rK^{-1}
\psi_r=\lambda\psi_r$  and explicit formulas \r{negative} for $i=1,\ldots,r$
that follows from $\Psi=G(\Lambda)\Psio$ (see details in
\cite{DLPPOS,LP}).

Let us  lift up the discrete evolution \r{e**} to the $n\times n$  matrix form.
It is easy exercise to show using formulas \r{negative} that absence of
negative powers of $\lambda$  in the operator
\footnote{Subscripts $\scriptstyle-$ and $\scriptstyle+$ in \r{edisc} and
\r{ed1} mean the projections on the negative and nonnegative powers
of the spectral parameter $\lambda$.}
\beq\label{edisc}
\left[G(t_0^{(a)}+1)(\Lambda_r\otimes E_a)G(t_0^{(a)})^{-1}\right]_-=0
\eeq
is equivalent to the discrete evolution of the matrix pseudo differential
operator $K(x,\partial)$ \r{e*}.
The equation \r{edisc} shows that dressed solution
to the linear problems \r{lin1} and \r{lin2} satisfies also the discrete
equation
\beq\label{ed1}
\Psi(t_0^{(a)}+1)=
\left[G(t_0^{(a)}+1)(\Lambda_r\otimes
E_a)G(t_0^{(a)})^{-1}\right]_+\Psi(t_0^{(a)}) \eeq
and  $p$-component modified
KdV$_r$ hierarchy can be obtained as compatibility condition of the linear
problems \r{ed1} and \r{lin2}. On the other hand the compatibility condition of
the linear problems \r{ed1} and \r{lin1} yields the generalization of the Toda
chain hierarchy. Indeed, we will see in the next subsection that this
compatibility condition give exactly Toda chain hierarchy in
the limit case $r=1$.
In fact, the $p$-component modified KdV$_r$ hierarchy which follows from
\r{ed1}
and \r{lin2} has inconvenient form, because our  special choice of the dressing
 operator $G$. By this choice the  fields matrix $q^\c$ in the operator
$G(\Lambda)(\partial_x-\Lambda)G^{-1} (\Lambda)$  has the form \r{e11}
rather than block diagonal form and the mKdV$_r$ fields $m_a$
enter to the operator
$\left[G(t_0^{(a)}+1)(\Lambda_r\otimes E_a)G(t_0^{(a)})^{-1}\right]_+$
in a very complicated and nonlinear way.

To write down the modified hierarchy in more convenient form we introduce
new dressing operator $\tilde G(\Lambda)$ as the formal $n\times n$ matrix
series with respect to the spectral parameter
\beq\label{newser}
\tilde G(\Lambda)=\sum_{i=0}^{\infty}\tilde G_i\Lambda^{-i},\quad
\left(\tilde G_i\right)_{ab}=\left(\begin{array}{cccc}
H_{1}&&&\\ &H_{2}&&\\ &&\ddots&\\ &&&H_{r} \end{array}\right),
\eeq
where
\beq\label{ediag1}
H_{i}=\sum_{a=1}^{p}K_i(t^{(a)}_0+r-i)E_a
\eeq
and $K_i$ and $H_{i}$ are $p\times p$  matrix functions. The
theorem follows

\theor{6}{The
$p$-component modified KdV$_r$ hierarchy  is defined by
the compatibility condition
of the   linear problems for the column  $\tilde\Psi=\tilde
G(\Lambda)\Psio$
\beq\label{ediag2}
(\partial-\Lambda+q^{\rm diag})\tilde \Psi=0,
\eeq
\beq\label{ediag3}
(\partial_{t^{(a)}_s}-\tilde V(\lambda))\tilde\Psi=0.
\eeq
The discrete evolution
\beq\label{ediag4}
\tilde\Psi(t_0^{(a)}+1)=
\left[\tilde G(t_0^{(a)}+1)(\Lambda_r\otimes
E_a)\tilde G(t_0^{(a)})^{-1}\right]_+\tilde\Psi(t_0^{(a)})
\eeq
is trivially satisfied due to \r{e32} and
block diagonal matrix $q^{\rm diag}$ is
\beq\label{ediag5}
q^{\rm diag}={\rm diag}\,(M_r,\ldots,M_1), \quad{\rm where}\quad
 M_i=\Delta^{-1}\sum_{a=1}^{p}m_a(t^{(a)}_0+i-1).
\eeq
The matrix $\tilde V(\lambda)$ differs from the matrix $V(\lambda)$ by the
gauge transformation  which does not depend on the spectral parameter
$\lambda$
\beq\label{ediag6}
\tilde V(\lambda)=S_{t^{(a)}_s}S^{-1}+SV(\lambda)S^{-1},
\eeq
where the $p\times p$ blocks of the matrix $S$ are defined by
$(1\leq a,b\leq r)$
\beq\label{ediag7}
S_{ab}={\rm res}\left[\left(\sum_{c=1}^{p}
((\Delta\partial)E_c+m_{c}(t_0^{(c)}+r-a-1)
\cdots
((\Delta\partial)E_c+m_{c}(t_0^{(c)}))
\right)\partial^{-r-1+b}\right].
\eeq               }

\proof\ The proof is the direct calculation where the property
of the matrices $E_a$
\beq\label{eprop}
E_aE_b=\delta_{ab}E_a
\eeq
is used \cite{LP91}.
Since the gauge transformation \r{ediag6} does not depend on the
spectral parameter the coefficient of $\tilde V(\lambda)$ at $\lambda^0$
can be easily obtained from \r{ediag6}.

\subsection{Exceptional case of $r=1$}

In first section  we have defined the regular elements of the loop
algebra $\widetilde{sl}_n$ for the equal block partition excluding the case
$r=1$ (pure homogeneous picture). In this subsection we will show that this
case can be naturally included in
Zakharov-Shabat dressing construction.
The results presented here are well know from the early
years of the development of the soliton theory (see \cite{New} and reference
therein) and we will present these results here for reader's convenience.
The dressing
condition \r{e23} in this case is
\beq\label{e38}
L= K(\Delta\partial)K^{-1}
=\Delta\partial+u,\quad {\rm diag}\ u=0
\eeq
and can be solved for the operator $K$ due to $\Delta_i\neq
\Delta_j$. The
 time evolution of the matrix $u$ with respect to the continuum times $t^
{(a)}_s$ and discrete times $t_0^{(a)}$ are
 defined by
compatibility conditions of the following linear problems
\beq\label{linn1}
L\psi=(K(\Delta\partial)K^{-1})\psi=(\Delta\partial+u)\psi=\lambda\psi,
\eeq
\beq\label{linn2}
{\partial\psi\over\partial t_s^{(a)}}=(KE_a(\Delta\partial)^sK^{-1})_+\psi,
\eeq
\beq\label{linn3}
\psi(t_0^{(a)}+1)=
(K(t_0^{(a)}+1)E_a(\Delta\partial)K^{-1})_+\psi=(E_a(\Delta\partial)
+m_a)\psi.
\eeq
The compatibility condition of \r{linn1} and \r{linn2} can be written in the
form
\beq\label{e37}
{\partial
u_{ik}\over\partial t_s^{(a)}}=-\Delta_i\Delta_k
\left(X^{(1)}_{ik}\right)'+\sum_{j=1}^{p}
\left[\Delta_iX^{(1)}_{ij}u_{jk}-u_{ij}X^{(1)}_{jk}\Delta_k\right],
\eeq
where  matrix $X^{(1)}$ is defined by
\beq\label{oper}
X=\left(K(\Delta\partial)^{s-1}E_aK^{-1}\right)_-=
\sum_{j=1}^{\infty} \partial^{-j}\circ  X^{(j)}.
\eeq
This hierarchy  was called $p$-component NLS hierarchy in \cite{KvL93}.
The equations \r{e29b} allows one to reconstruct the Hamiltonians
corresponding to the hierarchy \r{e37} using
 $X^{(1)}_{ij}=\delta {\cal H}_{s,a}/\delta u_{ij}$.
It can be shown that diagonalization procedure \r{e21} by
means of the pseudo-differential operator $g$  yields the same
Hamiltonians as diagonalization procedure due to Drinfeld and Sokolov
\cite{DS} or the resolvent method introduced by Dickey \cite{Dibook}.
The hierarchies of the type \r{e37} were investigated in \cite{Di} for
the case of arbitrary $p$  and in \cite{BtK88} for the case  $p=2$.

When $s=1$ we obtain   equations
\bea\label{e39}
{\partial u_{ik}\over\partial t_1^{(a)}}&=&{\Delta_a(\Delta_i-\Delta_k)
\over (\Delta_a-\Delta_i)(\Delta_a-\Delta_k)}u_{ia}u_{ak},\quad a\neq i,k,
\nn\\
{\partial u_{ik}\over\partial t_1^{(i)}}&=&{\Delta_i\Delta_k
\over \Delta_k-\Delta_i}u'_{ik}-\sum_{j\neq i}{\Delta_i
\over \Delta_j-\Delta_i}u_{ij}u_{jk},\quad a= i,
\nn\\
{\partial u_{ik}\over\partial t_1^{(k)}}&=&-{\Delta_i\Delta_k
\over \Delta_k-\Delta_i}u'_{ik}+\sum_{j\neq k}{\Delta_k
\over \Delta_j-\Delta_k}u_{ij}u_{jk},\quad a= k,
\eea
which  become trivial in the case $p=2$.

For $s\geq 2$ the explicit form of the equation \r{e37} becomes more
complicated. Let us write down these equation for the simplest
case $p=2$. To do this we introduce new time variable $t_2$ (instead
dependent due to \r{e31c} times $t^ {(a)}_2$, $a=1,2$) such that
\beq\label{e40}
{\partial L\over\partial t_{2}}={1\over i}\left(
{\partial L\over\partial t_{2}^{(1)}}  -
{\partial L\over\partial t_{2}^{(2)}}\right)
\eeq
and set $\Delta_1=-\Delta_2=1$ for simplicity.
In terms of this variable the equations \r{e37} have most simplest form
$$
i{\partial u_{12}\over\partial t_2}= -{1\over2}u''_{12}-u^2_{12}u_{21},
\quad
i{\partial u_{21}\over\partial t_2}= {1\over2}u''_{21}+u_{12}u^2_{21},
$$
and coincide with NLS equation after identification $u_{21}=
\kappa\overline{u}_{12}$, where $\overline{u}_{12}$
means the complex conjugation
and $\kappa=\pm1$.

The compatibility condition of \r{linn1} and \r{linn3} yields the
multi-component analog of the Toda chain hierarchy. To see it, let us consider
explicitly two-component case. Due to the relation
\r{e***}
we can restrict ourself only to the evolution with respect to
$t_0^{(1)}$. Now it is easy exercise to show that relation
\beq\label{eToda}
(\Delta\partial+u(t_0^{(1)}+1))(E_1\Delta\partial
+m_1(t_0^{(1)}))=(E_1\Delta\partial+m_1(t_0^{(1)}))
(\Delta\partial+u(t_0^{(1)}))
\eeq
is equivalent to the first equation in the Toda chain hierarchy
\beq\label{eToda1}
\Delta_1\Delta_2{\partial^2\over
\partial x^2}\varphi(t_0^{(1)})+e^{\varphi(t_0^{(1)}+1)-\varphi(t_0^{(1)})}-
e^{\varphi(t_0^{(1)})-\varphi(t_0^{(1)}-1)}=0
\eeq
and elements of the matrix $u$ appear to be connected at the different
values of the discrete time $t_0^{(1)}$
\beq\label{eToda2}
u_{12}(t_0^{(1)})=u_{21}(t_0^{(1)}+1)^{-1}=e^\varphi(t_0^{(1)}).
\eeq

\section{Conclusion}

We have shown that Zakharov-Shabat dressing technique can be applied
for the generalized integrable hierarchies discussed recently in
\cite{dGHM,FHM}. It allows one to conclude that there are exist
single dressing formalism that can be used for the description both
principal
(generalized KdV)  and homogeneous (generalized  AKNS) hierarchies as
well as all intermediate ones.

To conclude let us outline the main steps of the construction.
First, we started with a Hamiltonian reduction for the loop Lie algebra
$\tilde g$ admitting the regular element in the Heisenberg subalgebra.
Solving the tangency constraints we have found an explicit expression
for the gradient of the gauge invariant functional defined on the
constrained manifold such that the corresponding
Hamiltonian vector field ensures a tangency condition.
Second, we  developed a dressing technique in terms of
pseudo-differential operators and formal series with respect to the
spectral parameter and proved the coincidence of the  integrable
hierarchies which appeared in the framework of dressing approach
with integrable  hierarchies which emerged after Hamiltonian reduction
on the reduced manifold $\overline M$.

All calculations have been carried out for the simplest case of the
loop Lie algebra $\widetilde{sl}_n$. It will be interesting to repeat
these explicit calculations for the $BCD$ and exceptional Lie algebras.
In this case the first step of the construction can be easily repeated
using ideas of the seminal paper \cite{DS}. To realize the second
step one has to use the classification of the Heisenberg subalgebras
admitting the regular elements
which was obtained recently in \cite{DF94}.
It will be also interesting to investigate the relation between
Hamiltonian structures of the $BCD$ generalized
Drinfeld-Sokolov hierarchies and  the ${\cal W}$-algebras.
These relations in case of affine Lie algebra
$\widehat{sl}_n$ were considered in \cite{FGMG94}. The work in
this direction is in progress.

\section{Acknowledgments}

The second author (S.P.) would like to acknowledge
Departamento de F\'\i sica
Te\'orica of Universidad de Zaragoza where the work has been completed.
This work   was partially supported by Direcci\'on General de
Investigaci\'on Cient\'\i fica y T\'ecnica (Madrid).

\end{document}